\documentclass{iau} 
\usepackage{graphicx}

\title[PN distances from Gaia] 
{Galactic planetary nebulae as absolute probes: The view from Gaia }

\author[Letizia Stanghellini et al.]   
{Letizia Stanghellini$^1$,
 Beatrice Bucciarelli$^2$, Mario G. Lattanzi$^2$, and Roberto Morbidelli $^2$}

\affiliation{$^1$National Optical Astronomy Observatory \\ 950 N. Cherry Avenue, Tucson (AZ) 85719, USA \\ email: {\tt lstanghellini@noao.edu} \\[\affilskip]
$^2$INAF-Osservatorio Astronomico di Torino (Italy) \\ email: {\tt lattanzi@oato.inaf.it, bucciarelli@oato.inaf.it, morbidelli@oato.inaf.it }}

\pubyear{2016}
\volume{323} 
\jname{Planetary nebulae: Multi-wavelength probes of stellar and galactic evolution}
\editors{Xiaowei Liu, Letizia Stanghellini, and Amanda Karakas, eds.}
\begin{document}

\maketitle

\begin{abstract}
We searched the first Gaia data release for Galactic central stars of planetary nebulae (CSPNe) for parallaxes in order to determine the distances of the hosting PNe. For the small sample of PNe for which a comparison is available, we show that distances derived from Gaia parallaxes agree, within the uncertainties, with the individual PN distances derived by other reliable methods. While Gaia parallaxes available for Galactic CSPNe are still few, and with high uncertainties, we studied the possibility of building a PN distance scale by using the Gaia distances as calibrators. We found that a scale built on the relation between the linear nebular radius and its surface brightness has promising future applications.

\end{abstract}

\section{Motivation}
Distances are among the most important basic parameters needed to study the physics of planetary nebulae (PNe) and their central stars (CSs). Large sample of Galactic PNe with reliable individual distances are sought for in a variety of applications. 
Distances to PNe have been derived from spectroscopic binary CSs, cluster membership, reddening, and nebular expansion, in $\sim$ 40 Galactic PNe (see Stanghellini et al. 2008, and references therein), which represents a small fraction of the sample of several hundred, spectroscopy-confirmed Galactic PNe (Frew et al. 2016). Statistical distances, calibrated on these known distances, are commonly used. The problem with some of the statistical distance calibrators is that they rely at some level on modeling and assumptions. For example, reddening distances assume that the interstellar absorption toward a PN is similar to that of nearby stars, and does not account for possible patchiness in the ISM, while expansion distances assume that PN ejecta evolve homogeneously without acceleration. As a consequence, both the individual distances and the distance scale derived based on those can be severely misleading. The best individual distances are those derived for spectroscopic binary CSs, but only few of those are available to date.

\begin{figure}
\begin{center}
 \includegraphics[width=4.3in]{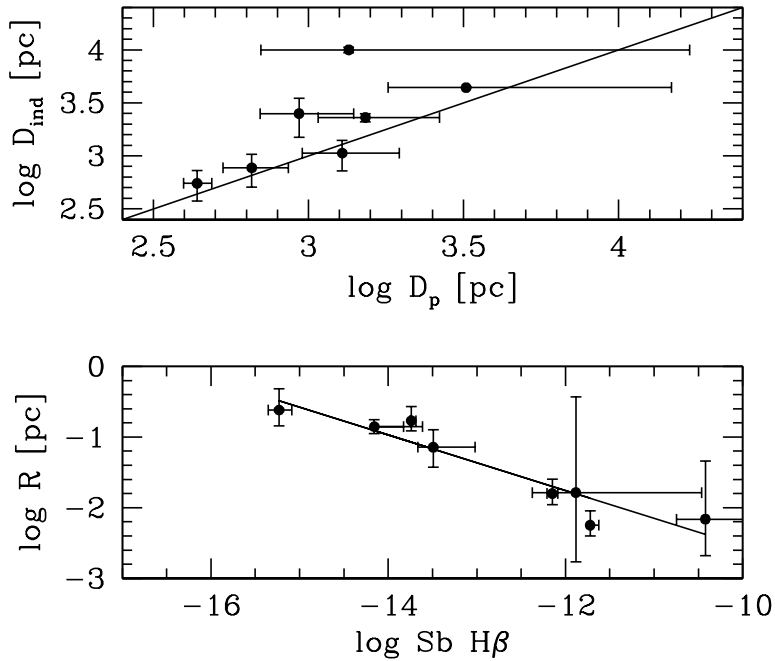} 
 \caption{Top panel: Comparison between Gaia parallax distances and independent distances, as given in Table 1. Bottom panel: The physical radius -- surface brightness relation obtained by calibrating with Gaia parallaxes. The resulting linear fit of the data, log R [pc] = -0.4 log Sb H$\beta$ - 6.5, is an initial guess of the Galactic PN distance scale derived from the first Gaia release. Excluding Sa~St~2-12 from the fit does not vary the correlation significantly.}
 \end{center}
 \end{figure}

\section{Parallaxes of CSPNe in the first Gaia data release}

The first Gaia data release (DR1, Gaia Collaboration et al. 2016) a few weeks ago has allowed us to retrieve the CSPN parallaxes that have been measured by Gaia with sufficiently low uncertainties to allow a distance determination. By searching DR1 against the astrometric positions corresponding to spectroscopically confirmed Galactic PNe (Acker et al. 1992, and updates) we found parallaxes for 8 CSPNe whose uncertainties are below 100$\%$, thus usable for a positive distance determination.  In Table 1 we list these targets (Cols. 1 and 2), the measured parallaxes of their CSs (3), their distances derived from Gaia parallaxes (4), and, if available in the literature, the independent distances (5), with the reference coded in Column (6).  Independent distances are from spectroscopic parallaxes, except for M~1-77 and NGC~2346, whose distances are derived with the extinction method. Note that the distance to Su~Wt~2 may refer to a companion star rather than the CS. It is also worth noting that the PN nature of Sa~St~2-12 may be uncertain. The high relative error of trigonometric parallaxes and the non-linear relations among the physical parameters forbid a direct application of the first-order error propagation formula; left and right (asymmetric) error bars are therefore estimated by inserting $x\pm\sigma(x)$ into the corresponding functional relation, $\sigma(x)$ being the known standard deviation of the observed quantity $x$.
Gaia parallax distances of CSPN span a broad range, and they agree with other independent distances within the errors, as shown in Fig.~1, where we plotted the available independent distances for the Gaia PN set against the best individual distances available in the literature. The best match is for nearby PNe, as expected, but even for PC~11, for which CS spectroscopy sets it at 10,000 pc, the two distances agree within the large parallax uncertainties. Note that all errorbars of the independent distances do not account for the intrinsic method uncertainties, which are typically very hard to quantify, as discussed in the original references listed in Table 1.

\begin{table}
  \caption{Distance of Galactic PNe from Gaia parallaxes and from other methods.}
  \scriptsize{
  \begin{tabular}{llrrrr}
  \hline 
{\bf Name} & {\bf Alias} & {\bf p$\pm\sigma$(p)} & {\bf log(D$_{\rm p}$)}&   {\bf log(D$_{\rm ind}$)} &ref\\ 
&&[mas]&[pc]&[pc]&\\
PN G038.2+12.0&       Cn~3-1&       1.932$\pm$0.656&  2.71$^{+0.18}_{-0.13}$&$\dots$&$\dots$\\    
PN G089.3-02.2&       M~1-77&       1.072$\pm$0.357&   2.97$^{+0.18}_{-0.13}$&3.025$^{+0.15}_{-0.22}$&HW88\\     
PN G165.5-15.2&     NGC~1514&       2.286$\pm$0.239&   2.64$^{+0.05}_{-0.04}$&  2.74$^{+0.12}_{-0.17}$&A15\\      
PN G215.6+03.6&     NGC~2346&       0.778$\pm$0.269&   3.11$^{+0.18}_{-0.13}$&3.025$^{+0.12}_{-0.17}$&G86\\   
PN G272.1+12.3&     NGC~3132&       1.524$\pm$0.364&   2.82$^{+0.12}_{-0.09}$& 2.89$^{+0.13}_{-0.18}$&C99\\
PN G311.0+02.4&       Su~Wt~2&      0.655$\pm$0.277&   3.18$^{+0.24}_{-0.15}$& 3.36$\pm$0.04&E10 \\
PN G331.1-05.7&        PC~11&      0.741$\pm$0.682&    3.13$^{+1.10}_{-0.28}$& 4.00$\pm$0.03&P10\\ 
PN G334.8-07.4&    Sa~St~2-12&      0.310$\pm$0.243&   3.51$^{+0.66}_{-0.25}$& 3.64&P04\\
     
  \hline 
    \end{tabular}
    }
  
\end{table}

\section{The Galactic PN distance scale from Gaia parallaxes}

While there are still too few calibrators, and their uncertainties remain relatively high, the relation between physical radius and surface brightness is well defined (see Fig. 1, bottom panel) at all surface brightness in our range. We use the surface brightness of PNe measured in H$\beta$, after correcting for the interstellar extinction. The linear correlation coefficient of the two sets of parameters of Fig. 1 (bottom panel) is 0.95, showing  a very tight correlation between the distance-dependent vs. the distance-independent parameter, make it a very promising distance scale. 
It is worth recalling that future data releases from Gaia will provide parallaxes of stars with V$<$15 mag with a precision of 0.03 mas (Lindegren et al. 2016). From Acker et al. (1992) we find that there are $\sim$50 Galactic PNe in this magnitude range, and whose statistical distances are estimated to be smaller than 3000 pc (Stanghellini et al., 2008). For this group of PNe, Gaia will provide final parallaxes of better than 10$\%$ relative uncertainties. For another $\sim$40 PNe the estimated parallax relative uncertainties will be of the order of 20$\%$. This wealth of new data will definitely help to constraint the scale. The present work is preliminary to set the stage for these future data sets.

\end{document}